\documentclass[aip,apl,reprint,superscriptaddress]{revtex4-1}

\usepackage{graphicx}
\usepackage{bm}
\usepackage{amsmath}
\usepackage{amsfonts}
\usepackage{color}

\begin{document}


\title{Induced quantum dot probe for material characterization}

 \author{Yun-Pil Shim}
 \affiliation{Laboratory for Physical Sciences, College Park, Maryland 20740, USA}
 \affiliation{Department of Physics, University of Maryland, College Park, Maryland 20742, USA}
 \author{Rusko Ruskov}
 \affiliation{Laboratory for Physical Sciences, College Park, Maryland 20740, USA}
 \affiliation{Department of Physics, University of Maryland, College Park, Maryland 20742, USA}
 \author{Hilary M. Hurst}
 \affiliation{Laboratory for Physical Sciences, College Park, Maryland 20740, USA}
 \author{Charles Tahan}
 \email{charlie@tahan.com}
 \affiliation{Laboratory for Physical Sciences, College Park, Maryland 20740, USA}
 \date{\today}

\begin{abstract}
We propose a non-destructive means of characterizing a semiconductor wafer
via measuring parameters of an induced quantum dot on the material system of interest with a separate probe chip
that can also house the measurement circuitry.
We show that a single wire can create the dot, determine if an electron is present, and be used to measure critical device parameters.
Adding more wires enables more complicated (potentially multi-dot) systems and measurements. As one application for this concept we consider silicon
metal-oxide-semiconductor and silicon/silicon-germanium quantum dot qubits relevant to quantum computing and show how to
measure low-lying excited states (so-called ``valley'' states).
This approach provides an alternative method for characterization of parameters that are critical for
various semiconductor-based quantum dot devices without fabricating such devices.
\end{abstract}

\maketitle


Semiconductor heterostructures often serve as the substrate for many solid-state devices.
For quantum devices such as qubits, their quality depends crucially on the properties of these wafers.
Often, these qubit characterization parameters can only be ascertained by fabricating the device and measuring it at cryogenic temperatures.
Quantum dots (QDs) in silicon for quantum computing (QC) \cite{Silicon_Review_RMP2013} are a great example.
The indirect band-gap of silicon creates low-lying excited (valley)
states in the QD heterostructure; if the ``valley splitting''
is too small, initialization, readout and even gate operation of the qubits is impeded.
Optimizing the valley splitting of silicon QD qubits---in addition to other important parameters such as coherence time,
charge noise, etc.---is needed for the eventual construction of quantum computers, and is limited by the design-fabrication-test cycle time.

We propose a method of characterizing material properties using a separate probe chip that both creates
the dot(s) and measures them. This concept was inspired by the ion trap stylus approach \cite{maiwald_wineland_nphys2009,arrington_review2013}
where an ion qubit is trapped on a stylus-like tip that can be brought close to a material to characterize its properties,
and also by the scanning nitrogen-vacancy (NV) center tip which can be used to detect magnetic fields at nanoscale for imaging or couple to
spin qubits\cite{casola_yacoby_nrevmat2018}.
While these ideas involve putting a qubit on the scanning tip itself,
our scheme uses a separate {\em gate chip} to {\em induce} a qubit in the material structure under study,
then measure those material and qubit parameters of interest using the circuits on the gate chip.
Indeed, scanning tunneling microscope (STM) tips have already been used to create effective dots on the surface of
InAs\cite{wittneven_prl1998,dombrowski_prb1999} and, more recently, Si\cite{salfi_prx2018}, using tunneling to do spectroscopy.
Nondestructive characterization of embedded donor atoms in a semiconductor has also been demonstrated
using a scanning tip architecture\cite{bussmann_rudolph_nanotechnology2015,gramse_kolker_sciadv2017}.
Here, we induce the dot qubit within the material in an environment realistic to quantum computing and consider dispersive readout
for characterizing material and qubit properties.


\begin{figure}
  \includegraphics[width=\linewidth]{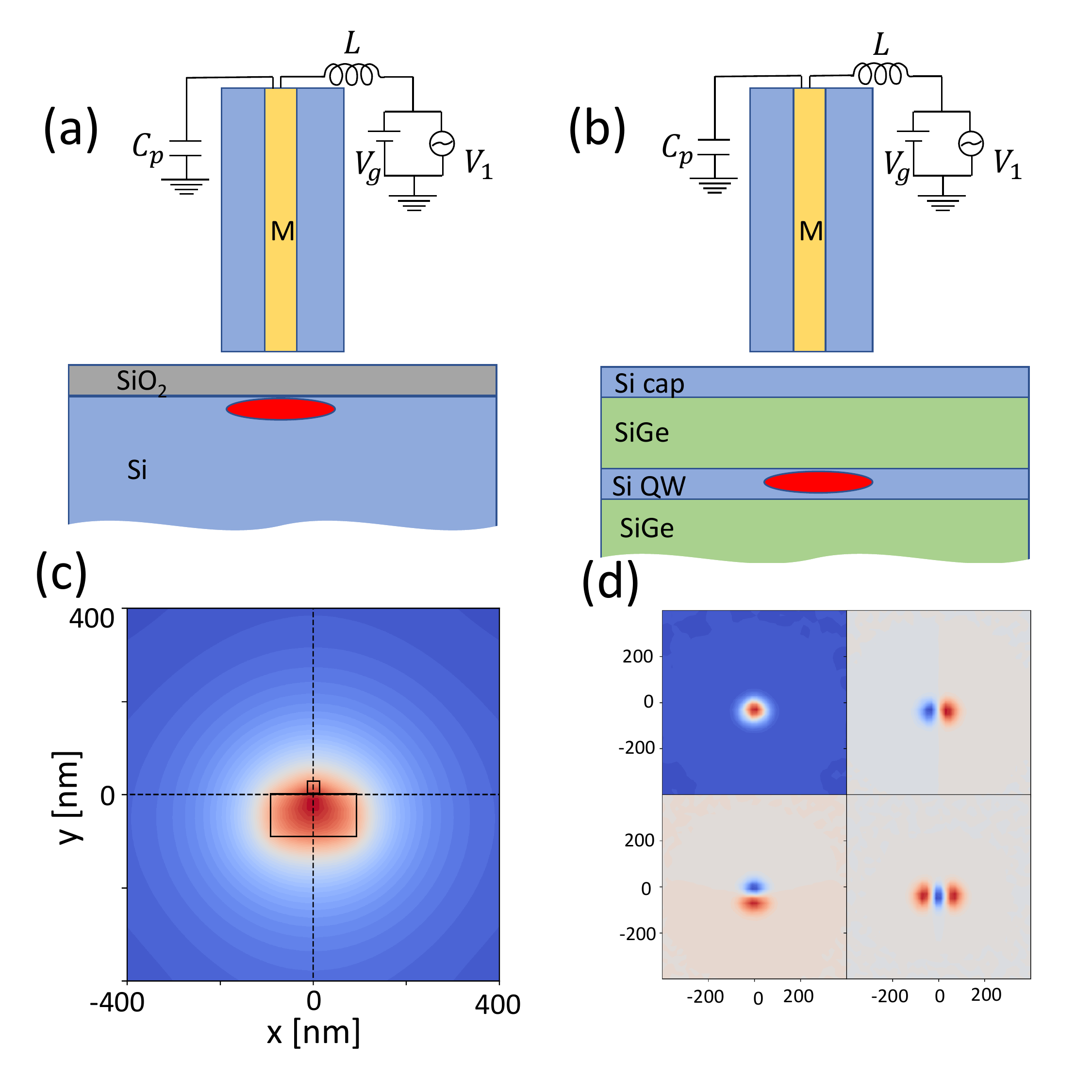}\\
  \caption{Schematic diagram of the device, for MOS (a) and QW (b) structures.
  A Si wafer chip with a metallic gate wire $M$ on it
  (and other necessary circuitry;
  $L$ is the inductance and $C_p$ is unavoidable parasitic capacitance)
  is positioned above the semiconductor heterostructure to induce a QD for
  non-invasive characterization.
  DC and AC voltages can be simultaneously applied to the gate wire for inducing the quantum dot and its characterization.
  Wire induced QD confining potential (c) and four lowest orbitals, ${\rm Re}[\psi_i]$, in a MOS device (d) with a DC gate voltage $V_g =0.02\,{\rm V}$.
}
  \label{fig:Fig1}
\end{figure}

To justify the viability of our approach we consider specifically silicon metal-oxide-semiconductor (MOS) type and
silicon/silicon-germanium quantum well (QW) type structures as examples to investigate relevant properties for
silicon-based qubit devices. We describe the general geometry of the heterostructure wafer and the gate chip and
provide electrostatic simulations of the induced QD.
Then, we show how to load the QD and detect the electron by
dispersive readout using the quantum capacitance of the induced QD all with the same wire.
Finally, methods for measuring the valley splitting based on a much stronger
quantum capacitance of the qubit levels at spin-valley anticrossing are discussed using one or more wires.

Figure \ref{fig:Fig1} shows schematic pictures of a possible setup.
The gate chip containing required
trapping and measurement circuitry is placed perpendicular above a semiconductor structure,
such as MOS [Fig. \ref{fig:Fig1} (a)] or Si/SiGe QW structure [Fig. \ref{fig:Fig1} (b)].
Applying positive voltage $V_g$ to the gate wire induces a confining electrostatic potential in the 2D quantum well in
the structure [Fig.~\ref{fig:Fig1} (c)] and orbital wave functions $\psi_i$
show typical 2D QD orbital characters [Fig.~\ref{fig:Fig1} (d)].
Electrons can be trapped into the induced QD as was depicted by red regions in Fig. \ref{fig:Fig1} (a) for MOS and (b) for QW.
The energy levels of the induced QDs have nonzero second derivative w.r.t. the applied voltage
(i.e. a quantum capacitance),
allowing for a dispersive readout by coupling to a detector circuit which can be integrated in the
gate chip \cite{colless_reilly_prl2013,zalba_barraud_ncomm2015,rossi_zhao_apl2017,urdampilleta_arxiv2018}.

We performed electrostatic simulation of the device\cite{COMSOL}
using dimensions for MOS and QW devices that are typically used in experiments.
For a MOS structure\cite{yang_rossi_ncomm2013,Veldhorst_Dzurak_nnano2014,Maurand_ncomm2016,Fogarty_arxiv2017},
a silicon oxide layer of 10nm overlays the silicon substrate of $\gtrsim 200\,{\rm nm}$.
For a QW structure\cite{kawakami_vandersypen_nnano2014,zajac_hazard_apl2015},
a strained silicon quantum well of 10nm is sandwiched between a $\gtrsim 200\,{\rm nm}$
SiGe substrate and a $40\,{\rm nm}$ SiGe spacer which is capped by 10nm of silicon.
We choose a reasonable and manufacturable gate chip design to demonstrate the main concepts in this work.
The gate wire size is chosen to be $10\,{\rm nm} \times 10\,{\rm nm}$ and $1\,\mu{\rm m}$ long,
and $10\,{\rm nm}$ away from the top of the heterostructure.
We considered different sizes of gate wafers as well as a bare metallic wire tip with no gate wafer for the
simulations and obtained qualitatively similar results. To be specific, we present below results for
the gate wire on a silicon wafer of
$100\,{\rm nm}$ depth and $200\,{\rm nm}$ width.

To conduct measurements of useful device properties, especially for properties relevant for spin qubits,
we need to populate the induced quantum dot with a controlled number of electrons.
This can be achieved in a number of different ways: e.g., (i) an electron-hole pair can be generated near the
induced QD by light, and the electron is trapped to the QD while the hole is pushed away from the QD by the electrostatic force,
or (ii) one can dope the semiconductor by implanting donors in a specific region
(or use large ``electron bath'' gate \cite{colless_reilly_prl2013,rossi_zhao_apl2017})
and use the dot accumulation wire to load electrons from the doped region into the QD
(one could then possibly move the electron to another area on the chip
as in the STM induced QD device \cite{salfi_prx2018}).
Once isolated, the dot gate voltage can be tuned to enhance the quantum capacitance while maintaining single occupation.


\begin{figure}
  \includegraphics[width=\linewidth]{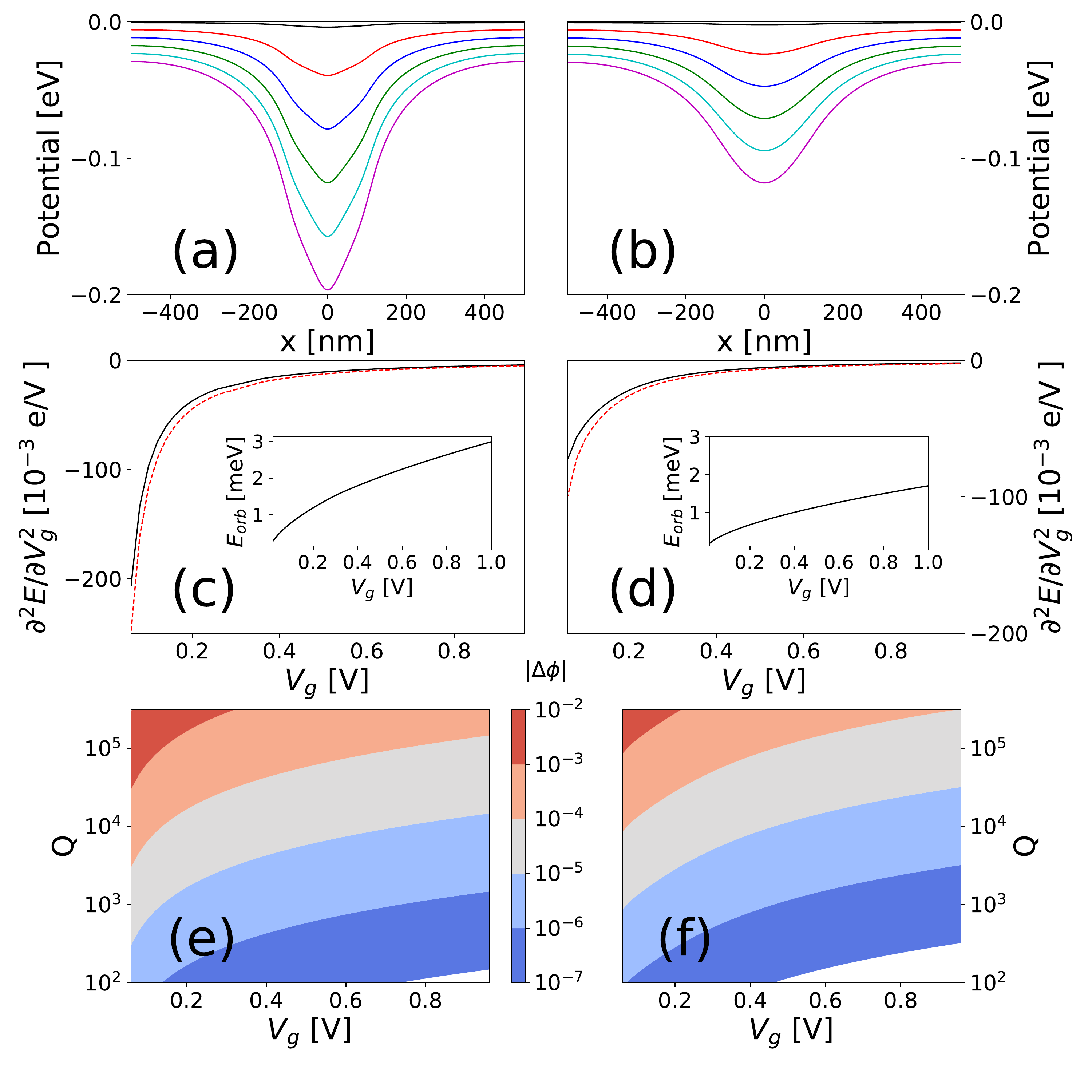}\\
  \caption{Simulation of the induced QD potential and
  energy levels' quantum capacitance, $\partial^2 E_i/\partial V_g^2$.
  (a) QD potential on a MOS device, for various gate voltages
$V_g=0.02,\, 0.2,\, 0.4,\, 0.6,\, 0.8,\, 1.0\,{\rm V}$ from top to bottom.
  (b) QD potential for a Si/SiGe QW device, for the same $V_g$ values.
  (c) and (d) are the quantum capacitances
  of the two lowest energy levels of the induced QD for MOS and QW structures, respectively.
  The solid black (dashed red) curves are for the ground (first excited) state orbitals.
  Insets show the energy splitting between the two lowest orbitals vs applied gate voltage $V_g$.
  (e) and (f) show the calculated phase shift, $\Delta \phi$, 
  of the reflected signal as a function of the voltage $V_g$ 
  and the quality factor $Q$ of the resonator circuit for MOS and QW devices, respectively, assuming the induced dot is singly occupied.}
  \label{fig:Fig2}
\end{figure}

We can detect the charge in the QD via dispersive
readout\cite{colless_reilly_prl2013,zalba_barraud_ncomm2015,rossi_zhao_apl2017,urdampilleta_arxiv2018}
by incorporating a tank-circuit (often superconducting) resonator (typically with frequency $\omega_r$ of
a few hundred MHz to a few GHz) into the gate wire and accumulated QD [e.g., Fig.~\ref{fig:Fig1}(a)],
and then sending and reflecting resonant microwaves to it.
There would be no phase shift of the reflected signal
from an empty dot, but if there is a trapped electron, the reflected signal will be
phase shifted if the quantum capacitance
\cite{AverinZorinLikharevZhETF1985,DutyDelsingPRL2005,sillanpaa_lehtinen_prl2005,petterson_smith_nl2010,CottetMoraKontos-prb2011,mizuta_otxoa_prb2017,RuskovTahan-PRL}
of the electron energy level is large enough.
We send an rf-signal (along with the DC voltage $V_g$):
$V = V_g + V_1 \cos(\omega_r t)$.
In addition to the conventional capacitance of the gate-to-heterostructure QD, $C_{\rm MOS}$,
and a distributed
parasitic capacitance $C_p$ of the gate to the ground plane, as was depicted in Fig.\ref{fig:Fig1} (a) and (b),
there will be a quantum capacitance $C_{\rm q,i} = \alpha_c^2 \frac{\partial^2 E_i}{\partial V_g^2}$
of the induced QD,
including the lever arm $\alpha_c \equiv \frac{C_c}{C_c + C_d}$ of the tip-to-dot capacitive coupling
(here $C_c$ and $C_d$ are the tip-to-dot and dot-to-ground capacitances, respectively;
for further estimations we assume $\alpha_c \sim 1$).
The quantum capacitance arises from the  non-linear voltage response of the QD's energy levels \cite{RuskovTahan-PRL},
$E_i(V_g) = E_i(V_g^0) + \frac{\partial E_i}{\partial V_g} \delta V_g + \frac{1}{2}\frac{\partial^2 E_i}{\partial V_g^2} \delta V_g^2$,
assuming slow in time voltage perturbation $\delta V_g$.
It leads to a frequency shift of the tank-circuit \cite{RuskovTahan-PRL},
and the corresponding phase shift of the reflected signal due to $C_{\rm q}$  would be \cite{colless_reilly_prl2013,zalba_barraud_ncomm2015}
\begin{equation}
\Delta \phi \simeq   Q\, \frac{\delta C}{C_{\rm tot}} \equiv  Q\, \frac{C_{\rm q}}{C_p + C_{\rm MOS} + C_q}  ,
\label{phase_response}
\end{equation}
where the Q-factor is defined via the tank-circuit relaxation $\kappa = \omega_r/Q$.
[For a single QD level the non-linear voltage response arises from the spatial change of the orbitals,
which is often neglected in Hubbard-like Hamiltonians. We recently showed the differences between a Hubbard-like
Hamiltonian and the actual induced QD system and its consequences on QD devices.\cite{ShimTahanPRB2018}]
Figure \ref{fig:Fig2} (a) and (b) show the QD confining potentials at various $V_g$ values for
MOS and QW devices, respectively. The second derivative of the orbital energy levels w.r.t the applied $V_g$
is shown in Fig. \ref{fig:Fig2} (c) for MOS and (d) for QW.
The absolute value of the quantum capacitance is larger for smaller gate voltage $V_g$
and can be as large as $\lesssim 0.03\, {\rm aF}$ for MOS
and $\lesssim 0.01\, {\rm aF}$ for QW
at $V_g = 0.02\, {\rm V}$, for the geometry studied in this work.

For typical low $Q$ tank-circuits\cite{colless_reilly_prl2013,rossi_zhao_apl2017}
having $C_{\rm tot}$ of a few hundred fF (and a frequency in the few hundred MHz range)
a capacitance change at level of a few aF is measurable\cite{rossi_zhao_apl2017},
leading to a phase shift $\Delta \phi \approx 10^{-4}-10^{-5}$.
Figure \ref{fig:Fig2} (e) and (f) show the calculated phase shift
for $C_{\rm tot} = 1000\, {\rm fF}$, vs. $Q$ and $V_g$ from Eq.~(\ref{phase_response})
(assuming $\frac{\delta C}{C_{\rm tot}} \ll 1/Q$).
The sensitivity  to measure a small quantum capacitance
will increase for moderately large tank-circuit Q-factors,
(e.g., the recently proposed high-kinetic inductance nano-wire resonators\cite{samkharadze-strong-photon2018}
with frequency of a few GHz and $Q \approx 10^3$ can be used in our proposed vertical gate circuit).
As an example, for $C_q \gtrsim  0.01 {\rm aF}$  as  per the simulation,
and a reachable resonator parameters\cite{samkharadze-strong-photon2018}:
$C_{\rm tot} \approx 30\, {\rm fF}$, $Q \approx 10^3$, one can obtain $\Delta \phi \gtrsim 3\,\times 10^{-4}$,
which is readily measurable\cite{zalba_barraud_ncomm2015,rossi_zhao_apl2017}.
The lowest detectable $C_q$ may be limited by unwanted variation in gate-to-QD capacitance
as a function of gate voltage (e.g., due to interface traps below the QD gate \cite{rossi_zhao_apl2017}).

If the device is in a configuration where the induced QD is close to an electron reservoir or another quantum dot,
then the charge stability diagram
can be mapped out directly using the tunneling capacitance,\cite{petterson_smith_nl2010}
where the response signal peaks at a charge transition
(similar to Ref. [\onlinecite{colless_reilly_prl2013}]).


\begin{figure}
  \includegraphics[width=\linewidth]{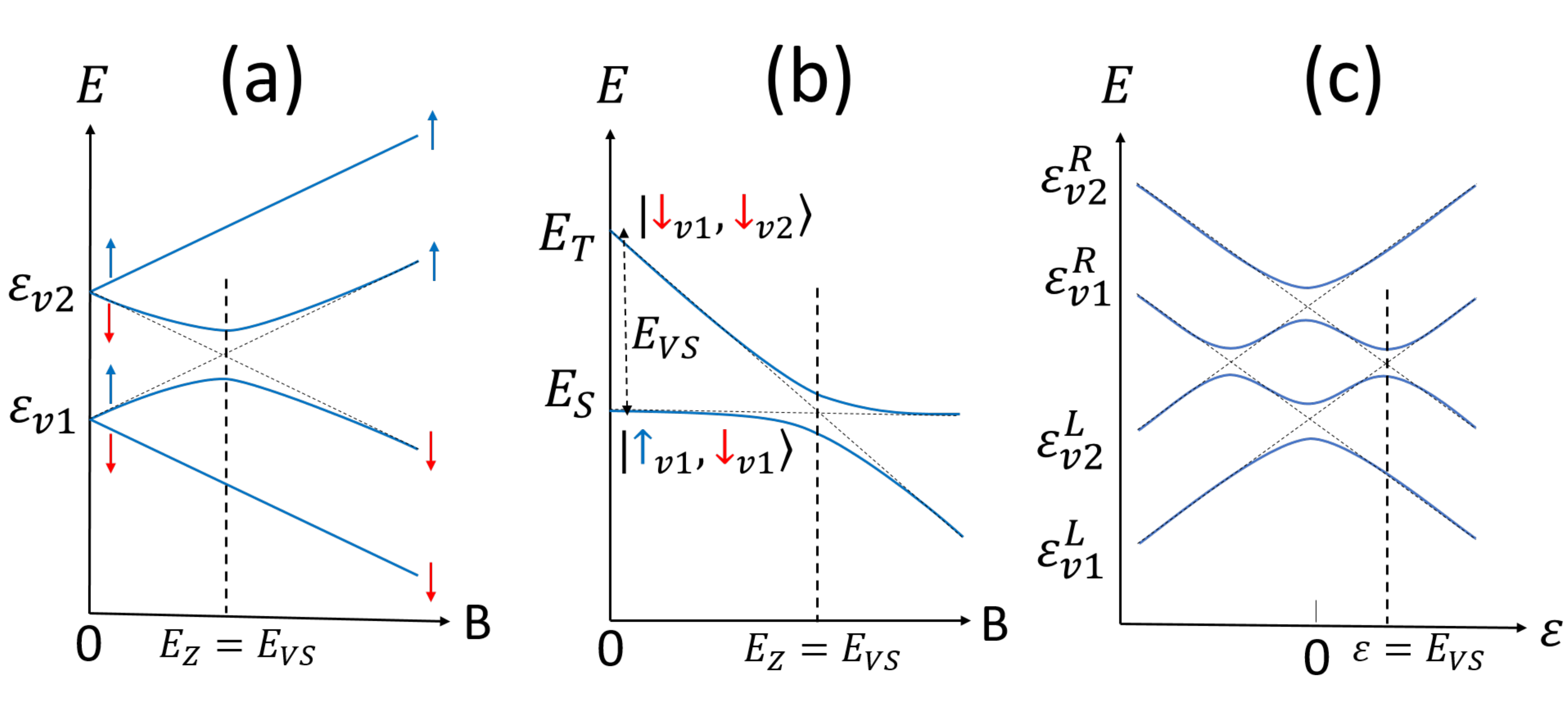}\\
  \caption{
  Schematic QD energy levels for valley-splitting measurement.
  (a) A QD with a single electron in an external magnetic field.
  The spin-valley states have a Zeeman splitting $E_{\rm Z}$,
  and when it is equal to the valley splitting $E_{\rm VS}$,
  there is an anti-crossing between the second and third levels. 
  The energy curvature w.r.t. gate voltage $V_g$ is maximal at the anticrossing [Eq.~(\ref{quantum-capacitance-B-field})]
  since $\frac{\partial^2 E_i}{\partial V_g^2} \propto \frac{\partial^2 E_i}{\partial B^2}$ for the regime considered.
  (b) A QD with two electrons in it has singlet and triplet states, which anti-cross at $E_{\rm Z}=E_{\rm VS}$.
  (c) A DQD with a single electron has anti-crossings vs dots' detuning $\varepsilon$, related to electron tunneling with (without)
  conservation of valley index.
  }
  \label{fig:Fig3}
\end{figure}


As an example of critical material parameters that the separate gate chip could measure,
we now discuss how to measure the valley splitting in a silicon wafer.
We first examine the case of a single QD with one electron.
Following the ideas of Ref.~[\onlinecite{yang_rossi_ncomm2013}], a relatively small accumulation mode QD as
in Ref.~[\onlinecite{rossi_zhao_apl2017}] can ensure that the orbital splitting is much larger than the valley splitting,
$E_{\rm orb} \gg E_{\rm VS}$,
which allows us to consider only the lowest orbital states in the following analysis.
Experimentally,
$E_{\rm VS} = 300 - 800\, \mu{\rm eV}$ and $E_{\rm orb} = 2-8\, {\rm meV}$ in small QDs in a MOS device\cite{yang_rossi_ncomm2013}
and also $E_{\rm VS} = 80 - 100\, \mu{\rm eV}$ and $E_{\rm orb} = 0.5\, {\rm meV}$ in another MOS device\cite{hao_ruskov_ncomm2014}.
For $\rm Si/SiGe$ quantum dots, $E_{\rm VS}$ could be of the order of $80 - 100\, \mu{\rm eV}$, or it could be much smaller.
In the simulation of induced dots the above is satisfied as $E_{\rm orb}$ is of the order of meV
[see insets of Fig.~\ref{fig:Fig2} (c) and (d)].

The valley splitting, $E_{\rm VS} \propto a V_g$, depends linearly on the applied top gate voltage\cite{yang_rossi_ncomm2013}.
By applying an in-plane magnetic field, the lowest two valley states are Zeeman split
(with energy splitting $E_{\rm Z}$) into 4 levels, as is shown in Fig.~\ref{fig:Fig3}(a).
The levels 2 and 3 (with different valley content) anti-cross when $E_{\rm Z} = E_{\rm VS}$,
which leads to levels' energy curvature w.r.t. the gate voltage $V_g$.
Indeed,
the splitting at anti-crossing, $\Delta_a \propto |\bm{r}_{v1,v2}| \, E_{\rm VS} \, (\beta_{D} - \alpha_{R})$
can be phenomenologically parameterized with an
(intervalley)
dipole matrix element $\bm{r}_{v1,v2}$ \cite{yang_rossi_ncomm2013,Gamble2013PRB},
implying a charge re-distribution as a result of interface-induced intervalley tunneling
and spin-orbit couplings\cite{RuskovVeldhorstDzurakTahan-PRB2018}.
We have estimated \cite{yang_rossi_ncomm2013}  $\Delta_a  =  10^{-4} - 10^{-3}\,  E_{\rm VS}$,
using a Rashba/Dresselhaus spin-orbit interactions, $\alpha_{R},\beta_{D}$, induced at the
heterostructure interface\cite{RuskovVeldhorstDzurakTahan-PRB2018}.
The levels 2 and 3 then read:
$E_{2,3}(V_g) = \frac{1}{2} [E_{\rm VS}(V_g) \mp \sqrt{(E_{\rm VS}(V_g) - E_{\rm Z}(B))^2 + \Delta_a(V_g)^2} ]$.
This was used to describe the relaxation ``hot spot''
observed in the experiment, which is mainly due to acoustic phonon emission\cite{yang_rossi_ncomm2013}.

Given this explicit level structure we calculate the curvature of the levels with respect to the
gate voltage $V_g$, obtaining the levels' quantum capacitances, $C_{\rm q,i}$
(this quantifies the non-linear response of the QD system\cite{RuskovTahan-PRL}).
In the magnetic field at anti-crossing [Fig.~\ref{fig:Fig3} (a)],
these quantum capacitances may be strongly enhanced w.r.t. that of the simple orbitals discussed above.
The ground state has zero curvature ($C_{\rm q,1}$=0) from this effect,
while for levels 2 and 3 one gets $C_{\rm q,2} = - C_{\rm q,3}$, and
\begin{equation}
C_{\rm q,3} \simeq \frac{a^2}{2\,\Delta_a} / \left[ \left( \frac{E_{\rm VS} - E_{\rm Z}}{\Delta_a} \right)^2 + 1 \right]^{3/2} ,
\label{quantum-capacitance-B-field}
\end{equation}
with the capacitances sharply peaked near the anti-crossing
(using a simple model with linear dependence on $V_g$ for the valley splitting\cite{RuskovVeldhorstDzurakTahan-PRB2018}).
With the experimentally estimated $\Delta_a$ and
valley splitting slope\cite{yang_rossi_ncomm2013} $a_{\rm exp} \simeq 0.64\, {\rm meV/V}$, we obtain
(for $E_{\rm VS} = 100\, \mu{\rm eV}$)
$|C_{\rm q,2,3}| \simeq 0.3 - 3 \, {\rm aF}$,
which should  be measurable in experiments\cite{zalba_barraud_ncomm2015,rossi_zhao_apl2017}.
Another capacitance contribution may appear due to fast relaxation processes\cite{mizuta_otxoa_prb2017}.
While the relaxation rate $\Gamma_{\rm rel}$  strongly increases at the spin-valley anti-crossing
for a single electron QD\cite{yang_rossi_ncomm2013} (reaching $10^7 - 10^8\, {\rm s}^{-1}$), it is much slower than the chosen
tank-circuit frequencies, $\Gamma_{\rm rel} \ll \omega_r$,
thus suppressing this capacitance contribution\cite{mizuta_otxoa_prb2017}.
A way to enhance $|C_{\rm q,2,3}|$ is to
use the in-plane magnetic field with an angle such that
$\Delta_a$ becomes much smaller\cite{RuskovVeldhorstDzurakTahan-PRB2018},
however making $\Delta_a$ smaller will  narrow the region
where $C_{\rm q}$ is significantly non-zero.

By scanning (sweeping) the magnetic field we will register a sharp peak of phase change of the
reflected signal when the Zeeman splitting is $E_{\rm Z} = E_{\rm VS}$.
For this to work, we need to populate the excited states by choosing a temperature comparable to the valley splitting,
e.g. for $E_{\rm VS} = 100\, \mu{\rm eV}$ the temperature should be $T \approx 1\, {\rm K}$.
Since $E_{VS} \gg \Delta_a$, the populations of the levels 2 and 3 in Fig.~\ref{fig:Fig3}(a) will be comparable,
thus leading to an effective quantum capacitance suppression by $\Delta_a/kT \sim \Delta_a/E_{\rm VS} \sim 10^{-3}$.

A way to mitigate these effects would be to use a single QD with  two electrons.
As shown in Fig.~\ref{fig:Fig3}(b), the lowest two levels now anti-cross at $E_{\rm Z} = E_{\rm VS}$
with an anti-crossing splitting $\Delta^{\rm 2e}_a  \approx  \Delta_a$ (scf. Ref.~[\onlinecite{yang_rossi_ncomm2013}]),
and the quantum capacitance is the same as in the 1-electron case,
while the relaxation is strongly suppressed at anti-crossing.
Also, the suppression effect due to temperature will not be as strong as in the 1-electron case,
since $kT \sim \omega_r \lesssim 1\,{\rm GHz}$ and so $\Delta_a/kT \sim \Delta_a/\omega_r \sim 10^{-1} - 10^{-2}$.
Since, however, we are in a regime $\omega_r \gg \Delta_a$
(opposite to that where a quantum capacitance approximation is valid) the effective quantum capacitance is suppressed by a form factor:
$C_{\rm q,eff} \simeq C_{\rm q} \, \left( \Delta_a / \omega_r\right)^2$. 
E.g., for $E_{\rm VS} \lesssim 100 \, \mu{\rm eV}$ the suppression factor is
$\left( \Delta_a / \omega_r\right)^2 \approx 1/40^2$.
Thus, this method would be sufficient to measure not too small valley splitting.


An alternative method to measure the valley splitting with a slightly more complicated gate circuit is
to induce a double QD using two or three gate wires on the gate chip.
Let us consider a DQD with a single electron, assuming each QD has the
same valley splitting.
The detuning between the QDs can be changed by tuning the voltages on the two QD-defining gates.
At zero detuning ($\varepsilon = 0$), one is at the degeneracy point of the lower eigenvalley $v1$-electrons.
[$v1$ is the lower valley and $v2$ is the upper valley states.
See Fig.~\ref{fig:Fig3}(c)].
The left-right tunneling $t$ between the dots defines the splitting at anti-crossing, $2 t$.
One then can measure the change of the reflected signal at the degeneracy point
(where the energy curvature is maximal) using a tank-circuit frequency $\omega_r \ll 2 t$.
By sweeping the detuning to $\varepsilon = E_{\rm VS}$ the $v1$-electron from the left can tunnel to the $v2$-level from the right.
This tunneling possibility forms another anti-crossing
and corresponding splitting (assume the same $2 t$).
(This kind of tunneling is briefly discussed in Ref.~[\onlinecite{hao_ruskov_ncomm2014}]
and then at length in Ref.~[\onlinecite{burkard_petta_prb2016}].)

To measure the valley splitting, one starts at $\varepsilon = 0$,
and populates the lowest two levels by temperature.
One then moves (faster than the relaxation time $T_1$) to a detuning  $\varepsilon = E_{\rm VS}$,
while sending a microwave with $\omega_r \ll 2 t$, to encounter a sharp change in the reflected phase (provided that $t \ll E_{\rm VS}$).
This can be fulfilled for $2 t \approx 2 - 4\, {\rm GHz}$ and $\omega_r \approx 0.5 - 1\, {\rm GHz}$.
Once $\varepsilon = E_{\rm VS}$ is reached, the reflected signal changes accordingly, due to maximal quantum capacitance
$C_{\rm q} =e^2/2t$ similar to the experiment of Pettersson et al.\cite{petterson_smith_nl2010}.
The quantum capacitance at this anti-crossing
is estimated of the order of $10\, {\rm fF}$,
which is several orders of magnitude larger than at the spin-valley anti-crossing discussed above.
In order to be able to distinguish the anti-crossings at $\varepsilon = 0$ and at $\varepsilon = E_{\rm VS}$,
one needs $E_{\rm VS} \gtrsim 2 t$ which sets the lowest measurable valley splitting, $E_{\rm VS} \gtrsim 5-10\, \mu{\rm eV}$.
The main difference of this proposal from that of Ref.~[\onlinecite{burkard_petta_prb2016}] is that the probing signal is far off
resonance with the level splitting, at a constant tank-circuit frequency $\omega_r \ll 2 t$, and the signature of
valley splitting is easier to measure.
These last two cases - doubly occupied single QD and a singly occupied DQD - may be the easiest to experimentally implement as a first verification of this proposed methodology.

Finally, we note that an additional (tunable) microwave field can be introduced to the above proposed experiments to drive transitions
between quantum dot states, which may allow for further or improved characterization (and also introduces another
absolute energy scale to compare to, in addition to the magnetic field).

The proposal presented in this paper requires
sensitive measurement of
small (quantum) capacitance changes, $C_q$, in sub-aF to aF range.
The signal due to $C_q$ may be obscured, however, in the presence of noise of
the large fluctuating capacitances, $C_{\rm MOS}$, $C_p$ (see Fig.~1).
For example, fluctuations via the voltage dependence of $C_{\rm MOS}(V_g)$  are attributed
to  the charging of interface traps below the QD gate \cite{rossi_zhao_apl2017};
the corresponding noise variation would be
$\Delta C_{\rm MOS}^{\rm noise} = |\frac{\partial C_{\rm MOS}}{\partial V_g}| \sqrt{\frac{S_V}{2t_{\rm av}}}$
where $S_V$ is the voltage spectral density and $t_{\rm av}$ is an averaging time.
It was  experimentally shown that below a sample-specific
voltage threshold, $V_g < V_{\rm th}$, the capacitance derivative is small
and a capacitance change of $\approx 1.5\, {\rm aF}$ was resolved \cite{rossi_zhao_apl2017}.
Another type of noise may enter through mechanical fluctuations of the tip. See e.g. experiments with
Scanning Microwave Microscopy (SMM) \cite{gramse_kolker_sciadv2017,deGraafDanilovAdamyanKubatkin-RevSciInst2013}.
In an experiment of near-field SMM \cite{deGraafDanilovAdamyanKubatkin-RevSciInst2013}
the (slow) resonator frequency fluctuations are tracked and stabilized via a feedback loop
allowing longer averaging time to reduce the noise;
a sensitivity of $(0.06\, {\rm aF})^2/{\rm Hz}$ was limited by mechanical noise\cite{deGraafDanilovAdamyanKubatkin-RevSciInst2013}.
Since in our  proposed experiment
the tip is not moving, the mechanical noise may be reduced,
eventually allowing for valley splitting measurement
via a tip-induced QD in the Si heterostructure.


Inducing quantum dots instead of fabricating them offers the potential for non-destructive characterization either locally or across a wafer,
speeding optimization of materials and quantum devices such as qubits.
Our concept is applicable to other materials and systems as the inducing and measurement chip
can be fabricated on a substrate different from the materials system under consideration.
We show that inducing QDs and measuring valley splitting in silicon devices is plausible
with current experimental technology.
Induced QD devices and the actual quantum devices built on the wafer will be different, but they share many critical aspects of
the underlying material. Characterization of the induced QD devices will provide useful information of the yet-to-be-built devices.
Based on this concept, other materials and systems (germanium, holes instead of electrons, topological systems, etc.) and qubit approaches
(encoded qubits, different readout techniques, even linear arrays of qubits making small quantum computers)
can be explored without actually fabricating the quantum dots themselves.

We thank Bob Butera and Michael Dreyer for helpful discussion.
H.M.H. acknowledges support from the NPSC fellowship.


\bibliographystyle{apsrev4-1}

%

\end{document}